\documentclass[11pt]{article}
\usepackage{axodraw}
\usepackage{epsfig}
\usepackage{amsfonts}
\usepackage{amsmath}
\usepackage{bbm}
\usepackage{cite}
\allowdisplaybreaks[1]

\input pix.sty

\newcommand{\T}{\rmii{$T$}}

\renewcommand{\eq}{eq.~}
\renewcommand{\eqs}{eqs.~}

\renewcommand{\se}{sec.~}

\renewcommand{\fig}{fig.~}

\newcommand{\mD}{m_\rmii{D}}

\newcommand{\Nf}{N_{\rm f}}
\newcommand{\Nc}{N_{\rm c}}

\newcommand{\Tc}{T_{\rm c}}

\newcommand{\rmO}{{\mathcal{O}}}

\newcommand{\CF}{C_\rmii{F}}

\def\lsi{\raise0.3ex\hbox{$<$\kern-0.75em\raise-1.1ex\hbox{$\sim$}}}
\def\gsi{\raise0.3ex\hbox{$>$\kern-0.75em\raise-1.1ex\hbox{$\sim$}}}
\newcommand{\lsim}{\mathop{\lsi}}
\newcommand{\gsim}{\mathop{\gsi}}

\newcommand{\rmii}[1]{{\mbox{\tiny\rm{#1}}}}

\newcommand{\re}{\mathop{\mbox{Re}}}
\newcommand{\im}{\mathop{\mbox{Im}}}

\newcommand{\Tint}[1]{{\hbox{$\sum$}\!\!\!\!\!\!\!\int\,}_{\!\!\!\!\raise-0.9ex\hbox{$\scriptstyle{#1}$}}}
\newcommand{\Tinti}[1]{{{\Sigma}\!\!\!\!\raise0.3ex\hbox{$\int$}_\rmii{${#1}$}}}

\newcommand{\bi}{\begin{itemize}}
\newcommand{\ei}{\end{itemize}}
\newcommand{\hide}[1]{ }

\newcommand{\deltabar}{\delta\!\!\!\raise0.7ex\hbox{--}\,}
\def\TAsc(#1,#2)(#3,#4,#5)%
{\SetWidth{2.0}\CArc(#1,#2)(#3,#4,#5)\SetWidth{1.0}}
\def\Lwidth{3}

\def\TAgl(#1,#2)(#3,#4,#5){\SetWidth{2.0}\PhotonArc(#1,#2)(#3,#4,#5){\Lwidth}%
{6.283 #3 mul 360 div #4 #5 sub #4 #5 sub mul sqrt mul Tdensity mul}%
\SetWidth{1.0}}
\def\TLgl(#1,#2)(#3,#4){\SetWidth{2.0}\Photon(#1,#2)(#3,#4){\Lwidth}
{#1 #3 sub #1 #3 sub mul #2 #4 sub #2 #4 sub mul add sqrt Tdensity mul}%
\SetWidth{1.0}}

\renewcommand{\picc}[1]{\;\parbox[c]{60pt}{\begin{picture}(60,30)(0,-3)
\SetWidth{1.0}\SetScale{1.1} #1 \end{picture}}\; }
\def\Lwidth{1.3}

\def\NRa{\picc{%
 \SetWidth{1.0} 
 \Line(0,15)(40,15)%
 \Line(40,15)(40,5)%
 \Line(40,5)(0,5)%
 \Line(0,5)(0,15)%
 \LongArrow(20,15)(22,15)%
 \LongArrow(40,10)(40,8)%
 \LongArrow(20,5)(18,5)%
 \LongArrow(0,10)(0,12)%
}}
\def\NRb{\picc{%
 \SetWidth{1.0} 
 \Line(0,15)(40,5)%
 \Line(40,5)(40,15)%
 \Line(40,15)(0,5)%
 \Line(0,5)(0,15)%
 \LongArrow(9,12.75)(11,12.25)%
 \LongArrow(40,10)(40,12)%
 \LongArrow(31,12.75)(29,12.25)%
 \LongArrow(0,10)(0,12)%
}}

\makeatletter \@addtoreset{equation}{section} \makeatother

\makeatletter
\renewcommand\section{\@startsection {section}{1}{\z@}%
                                   {-5.5ex \@plus -1ex \@minus -.2ex}
                                   {2.3ex \@plus.2ex}%
                                   {\normalfont\large\bfseries}}
\renewcommand\subsection{\@startsection{subsection}{2}{\z@}%
                                     {-3.25ex\@plus -1ex \@minus -.2ex}%
                                     {1.5ex \@plus .2ex}%
                                     {\normalfont\normalsize\bfseries}}
\renewcommand\thesection {\@arabic\c@section}
\renewcommand\thesubsection   {\thesection.\@arabic\c@subsection}
\renewcommand{\@seccntformat}[1]{%
\csname the#1\endcsname.\hspace{1.0em}}
\makeatother

\begin{document}

\flushbottom

\begin{titlepage}

\begin{flushright}
June 2019
\end{flushright} 
\begin{centering}

\vfill

{\Large{\bf
Studies of a thermally averaged $p$-wave Sommerfeld factor
}} 

\vspace{0.8cm}

Seyong Kim$^{\rm a,b}$ and 
M.~Laine$^{\rm b}$ 

\vspace{0.8cm}

$^\rmi{a}$%
{\em
Department of Physics, 
Sejong University, Seoul 143-747, South Korea\\}

\vspace*{0.3cm}

$^\rmi{b}$%
{\em
AEC, Institute for Theoretical Physics, 
University of Bern, \\ 
Sidlerstrasse 5, CH-3012 Bern, Switzerland\\} 

\vspace*{0.8cm}

\mbox{\bf Abstract}

\end{centering}

\vspace*{0.3cm}
 
\noindent
Thermal pair annihilation of heavy particles, such as dark matter or 
its co-annihilation partners, can be strongly influenced by attractive
interactions. We investigate the case that pair annihilation proceeds through
a velocity-suppressed $p$-wave operator, in the presence of an SU(3) gauge
force. Making use of a non-relativistic effective theory, the thermal average
of the pair-annihilation rate is estimated both through a resummed
perturbative computation and through lattice simulation, in the range 
$M/T \sim 10 ... 30$. Bound states contribute to the annihilation process 
and enhancement factors of up to $\sim 100$ can be found.

\vfill

 

\vfill

\end{titlepage}

%
\section{Introduction}
\la{se:intro}

Inelastic processes between a dilute ensemble of heavy particles moving slowly 
in a thermal environment are encountered in many physical situations. 
A classic example is given by nuclear reactions taking place 
within the electromagnetic plasma of stars~\cite{lsb}. 
In particle physics, we may consider heavy dark matter 
particles pair-annihilating into Standard Model particles 
in the early universe, or a heavy quark and anti-quark pair-annihilating 
into light quarks and gluons in a quark-gluon plasma generated in 
heavy ion collision experiments. 

The theoretical treatment of slow annihilation 
processes is facilitated by noting
that the average kinetic energy of the annihilating particles
is small compared with their rest mass, $Mv^2 \sim T \ll M$. 
Such a scale separation permits for a factorized description
of annihilation processes in terms of a series of long-distance matrix
elements times short-distance Wilson coefficients~\cite{bodwin}. 
In particular, the thermal average of an annihilation rate can 
be expanded as 
$
 \langle \sigma v \rangle = a + b\, \langle v^2 \rangle + \cdots \; 
$, where $v$ denotes the relative velocity. 
The term $a$ is said to originate from ``$s$-wave'' matrix elements, 
whereas $b$ may be associated with ``$p$-wave'' ones. 

In the presence of long-range interactions, the coefficients $a$ and $b$ may
get large corrections compared with a tree-level treatment. For scattering
states, this is known as  
the ``Sommerfeld (-Gamow-Sakharov) effect''~\cite{asommerfeld,landau3,gamow,sakharov}. 
Sommerfeld factors are nowadays routinely included in Boltzmann
equations for dark matter 
pair annihilation
(cf.,\ e.g.,\ refs.~\cite{hisano,sfeldx,feng,mb,etc}). 

Long-range interactions may also lead to 
the appearance of bound states in the 
dark sector, which opens up a fast pair-annihilation 
channel (cf.,\ e.g.,\ refs.~\cite{old32,old4}). 
Bound states are particularly important 
if the dark sector contains particles charged under QCD, as is the case
for instance in a prototypical model in which 
dark matter is a singlet Majorana fermion and the mediator is a
slightly heavier strongly coupled scalar
(cf.\ refs.~\cite{giv,mg3} for reviews).

Recently, we have developed a framework which permits to estimate
the thermally averaged pair annihilation rate, 
including bound-state effects, beyond perturbation 
theory~\cite{4quark_lattice}. The framework can be applied
to a number of cosmological models~\cite{threshold}, particularly
the prototypical framework mentioned above~\cite{stop,sb}, where
bound-state effects have been seen to be important from other
considerations as well~\cite{ll,mrss,klz,hp}.  

The purpose of the present work is to extend 
ref.~\cite{4quark_lattice} from the $s$-wave to the $p$-wave case. 
Even if the $p$-wave contribution is suppressed by $\langle v^2 \rangle$, 
its ``standard'' Sommerfeld enhancement is larger than for 
$s$-wave~\cite{iengo,cassel}.  
If the coefficients of the $s$-wave operators happen to vanish 
at leading order, $p$-wave may be the dominant channel~\cite{hg}. 
$p$-wave annihilation has also been discussed 
from astrophysical motivations 
(cf.,\ e.g.,\ refs.~\cite{Zhao:2016xie,Choquette:2016xsw,An:2016kie}).

This presentation is organized as follows. 
After outlining the basic setup 
(cf.\ \se\ref{se:basic}), we review 
thermally averaged pair annihilation rates
within resummed perturbation theory (cf.\  \se\ref{se:pert}). 
Having introduced the lattice framework 
(cf.\ \se\ref{se:latt}), we present and discuss numerical 
results (cf.\ \se\ref{se:num}), and conclude then
with a brief summary (cf.\ \se\ref{se:concl}). 

%
\section{Basic setup}
\la{se:basic}

Denoting by $n$ the dark matter number density, 
and assuming that there is a discrete quantum number 
which prohibits dark matter from decaying, its cosmological evolution is
normally described by the Lee-Weinberg equation~\cite{clas1,clas2,old1},
\be
 \dot{n} + 3 H n = - \langle \sigma v \rangle 
 \, \bigl( n^2 - n_\rmi{eq}^2 \bigr) 
 \;, \la{lw}
\ee
where $H$ is the Hubble rate, 
$\sigma$ is an annihilation cross section,
$v$ is a relative velocity, 
and 
$
 \langle ... \rangle
$
indicates a thermal average over the momenta of
the annihilating particles. 

If the dark sector experiences strong interactions, the thermal average 
$\langle \sigma v \rangle$ may receive large radiative corrections. 
In order to address these beyond perturbation theory, it was noted
in ref.~\cite{chemical} that by linearizing \eq\nr{lw} close to 
equilibrium, we may interpret the averaged cross section
as being related to a chemical equilibration rate 
($\equiv \Gamma^{ }_\rmi{chem}$), 
\be
 \langle \sigma v \rangle = 
 \frac{ \Gamma^{ }_\rmi{chem} }{ 2 n^{ }_\rmi{eq} }
 \;. 
\ee
Subsequently we can make use of linear response theory in order 
to relate $\Gamma^{ }_\rmi{chem}$ to an equilibrium correlator. 
Furthermore, if we find
ourselves in the non-relativistic regime, i.e.\ with dark matter 
masses $M \gg \pi T $, then the annihilations can be 
described by local operators~\cite{lsb}, similar to those found
in the NRQCD context~\cite{bodwin}. Then the equilibrium correlators
can be reduced to thermal expectation values of the annihilation
operators~\cite{4quark_lattice}, 
\be
 \langle \sigma v \rangle = 
 4 \sum_i c^{ }_i \,   
 \frac{\langle \rmO^{ }_i \rangle }{n_\rmi{eq}^2}
 \;. \la{c_i}
\ee
Here the Wilson coefficients $c^{ }_i$ and the operators
$\rmO^{ }_i$ can be taken over from a vacuum computation, capturing the
contribution of ``hard scales'' to the annihilation process, whereas
the influence of the ``soft scales'' resides within 
the thermal expectation value $\langle ... \rangle$. 

As is usual for effective field theories, 
the operators $\rmO^{ }_i$ can be organized as an expansion in $1/M^2$.
The leading terms, called $s$-wave operators, do not contain derivatives
and are suppressed by $1/M^2$. At the next order, operators appear
which contain two spatial derivatives and which are 
correspondingly suppressed by $1/M^4$. Given that 
$\langle \nabla^2 \rangle / M^2 \sim \pi T/M \ll 1$,
the $p$-wave operators are normally strongly suppressed compared
with the $s$-wave operators. However, $p$-wave operators may 
experience relatively speaking larger
enhancements from interactions~\cite{iengo,cassel}
and also display bound states, 
and they thus merit a detailed look.\footnote{%
 It has been suggested that, apart from influencing the value
 of $\langle \sigma v \rangle$, bound states also lead to a modification
 of the functional form of the part $n^2 - n_\rmi{eq}^2$ in 
 \eq\nr{lw} at late times when $n - n^{ }_\rmi{eq} \gg n^{ }_\rmi{eq}$
 so that we leave the linear response regime~\cite{binder}.
 Furthermore, when $\pi T \ll \Delta E$, where $\Delta E$ is a binding
 energy, bound states  
 fall out of chemical equilibrium, and should 
 be added as separate variables in the set of rate equations. 
 } 

The way that interactions modify the annihilation process can be 
parametrized through ``Sommerfeld factors''. In vacuum, the Sommerfeld
factor for an annihilation from unbound states is defined by writing
\be
 \sigma v = \sigma^{ }_\rmi{tree} v \times S(v)
 \;,
\ee
after which thermal averaging is often implemented as 
\be
 \langle \sigma v \rangle \;\simeq\; \frac{\int_\vc{v} \sigma v\, 
 e^{-M^{ }_\rmii{kin}v^2 / T}}
 {\int_\vc{v} e^{-M^{ }_\rmii{kin} v^2/T}} 
 \;. 
\ee
In reality, vacuum and thermal effects cannot be factorized in this way. 
Indeed thermal corrections can also modify masses like $M^{ }_\rmi{rest}$
and $M^{ }_\rmi{kin}$, and open 
up new channels not present in vacuum, like scatterings off
light plasma particles. 

A proper definition of a thermally averaged
Sommerfeld factor can be given for the combination appearing in 
\eq\nr{c_i} and for each operator separately, {\it viz.}\  
\be
 \bar{S}^{ }_i \;\equiv\; 
 \frac{ \langle \rmO^{ }_i \rangle /
        \langle \rmO^{ }_i \rangle^{ }_\rmi{tree}  }
      { n_\rmi{eq}^2 / 
       (n^2_\rmi{eq})^{ }_\rmi{tree} }
 \;. \la{Si_def}
\ee
Here we define 
$
 \langle \rmO^{ }_i \rangle^{ }_\rmi{tree}
$
and
$ 
 (n_\rmi{eq}^2)^{ }_\rmi{tree}
$
as tree-level quantities. The rationale of 
the double ratio in \eq\nr{Si_def} is that it 
removes effects not only from 
the tree-level scattering process but also from ``trivial'' 
corrections to the rest mass, which affect $n^{2}_\rmi{eq}$ and 
$
 \langle \rmO^{ }_i \rangle
$
by a large amount~\cite{lsb}.  
As a consequence of this definition, \eq\nr{c_i} can now be 
re-expressed as 
\be
 \langle \sigma v \rangle = 
 4 \sum_i c^{ }_i \, 
 \bar{S}^{ }_i \,   
 \frac{\langle \rmO^{ }_i \rangle^{ }_\rmi{tree} }
 {(n_\rmi{eq}^2)^{ }_\rmi{tree}} 
 \;, \la{c_i_new}
\ee
where the tree-level ratio 
$
 {\langle \rmO^{ }_i \rangle^{ }_\rmi{tree} } / 
 {(n_\rmi{eq}^2)^{ }_\rmi{tree}} 
$
is dimensionless and 
has a simple expression, 
for instance
as given in \eq\nr{ratio_tree} for the operator in \eq\nr{O_p}. 

To be concrete, we consider a theory with heavy particles charged 
under the fundamental and antifundamental representation of SU(3). 
Following the original inspiration from QCD~\cite{4quark_lattice}, 
these fields are taken to be a spin-$\fr12$ particle and antiparticle
(that is, heavy quark and antiquark), each with $N\equiv 2\Nc^{ }$ degrees
of freedom. However, spin-dependent effects are highly suppressed, 
so we believe our results to be valid also for spin-$0$ particles, 
such as stops, with the replacement $N\to \Nc^{ }$. 
The particle and antiparticle fields are denoted
by $\theta$ and $\chi$, respectively, and 
the annihilation operator considered
is defined in \eq\nr{O_p}.

%
\section{Perturbative considerations}
\la{se:pert}

Assuming that the overall scaling of the annihilation operators 
as $1/M^2$ has been incorporated into the coefficients $c^{ }_i$
in \eq\nr{c_i}, the $p$-wave operator that we consider is defined as   
\ba
 \mathcal{O}^{ }_p 
 & \equiv & 
 \frac{1}{M_\rmi{kin}^2} 
  \Bigl[\theta^\dagger \Bigl(
    - \frac{i}{2} \overleftrightarrow{D}
  \Bigr) \chi \Bigr] 
  \;
  \Bigl[\chi^\dagger \Bigl(
    - \frac{i}{2} \overleftrightarrow{D}
  \Bigr) \theta \Bigr]
 \;. \la{O_p}
\ea
Here $\theta$
and $\chi^\dagger$ are annihilation operators for particles and 
antiparticles, respectively. 
As the annihilation operators appear on the right,
the vacuum
state does not contribute to $\langle \rmO^{ }_p \rangle$.

It is straightforward to evaluate the thermal expectation 
value of \eq\nr{O_p} in tree-level perturbation theory. 
We obtain 
\be
 \bigl\langle \rmO^{ }_p
 \bigr\rangle^{ }_\rmi{tree} \; = \; N \int^{ }_{\vc{p,q}} 
 \frac{(\vc{p}-\vc{q})^2}{4 M_\rmi{kin}^2} e^{-(E^{ }_p + E^{ }_q)/T}
 \; = \; N \times \frac{ 3  T }{2 M^{ }_\rmi{kin}} \times 
 \biggl(\frac{M^{ }_\rmi{kin} T}{2\pi} \biggr)^3 
 e^{- {2 M^{ }_\rmi{rest} } / {T}} 
 \;,  \la{Op_tree}
\ee
where $E^{ }_p \equiv M^{ }_\rmi{rest} + p^2 / (2 M^{ }_\rmi{kin})$
is a non-relativistic energy.\footnote{%
 At $T > 0$, 
 $M^{ }_\rmii{rest}$ and $M^{ }_\rmii{kin}$ do not coincide because of the 
 so-called Salpeter correction to $M^{ }_\rmii{rest}$, 
 cf.,\ e.g.,\ refs.~\cite{lsb,chesler}. Even in vacuum the two can differ
 if UV regularization does not respect Lorentz 
 invariance, as is the case for instance within the lattice
 NRQCD setup. 
}
Similarly, 
\be
 \bigl( n^{ }_\rmi{eq} \bigr)^{ }_\rmi{tree} 
 \; \equiv \; 2N \int_{\vc{p}} e^{-E^{ }_p / T}
 = 
 2N\,
 \biggl(\frac{M^{ }_\rmi{kin} T}{2\pi} \biggr)^{\fr32} 
 e^{- {M^{ }_\rmi{rest} } / {T}} 
 \;, \la{neq}
\ee
and correspondingly 
\be
 \frac{ \langle \rmO^{ }_p
        \rangle^{ }_\rmi{tree} }
        { ( n^{2}_\rmi{eq} )^{ }_\rmi{tree} }
 \; = \; 
 \frac{ 3  T }{8 N M^{ }_\rmi{kin}}
 \;. \la{ratio_tree}
\ee
This displays a characteristic $p$-wave velocity suppression by 
$T/M^{ }_\rmi{kin}\ll 1$.

In order to determine the perturbative value of the averaged Sommerfeld 
factor of \eq\nr{Si_def}, it is helpful to go over into 
center-of-mass coordinates, defined as 
\be
 E^{ }_p + E^{ }_q
 \; = \; 
 2 M^{ }_\rmi{rest} + \frac{k^2}{4 M^{ }_\rmi{kin}} + E' 
 \;, \quad
 \vc{k} \; \equiv \; \vc{p+q}
 \;. \la{com}
\ee
Moreover, it is 
useful to resolve $\langle \rmO^{ }_p \rangle$ 
into a spectral representation, so that
contributions from soft energy scales can be inspected more carefully.
A thermal potential $V^{ }_\T(r)$ (cf.\ \eq\nr{imV})
is assumed normalized so that 
$\lim_{r\to \infty} V^{ }_\T(r) = 0$, i.e.\ $r$-independent thermal
corrections, known as the Salpeter correction, have been included
in the definition of $M^{ }_\rmi{rest}$. 
A vector-like Green's function is solved for from   
\ba
 \biggl\{
  -\frac{\nabla_\vc{r}^2}{M^{ }_\rmi{kin}}
 + V^{ }_\T (r) - i \Gamma^{ }_\T(r)  - E'   
 \biggr\} 
 \,\vc{G}(E';\vc{r},\vc{r}') & = & 
 \frac{N
   \, \nabla^{ }_{\vc{r}'} \delta^{(3)}(\vc{r} - \vc{r}')
 }{M_\rmi{kin}^2} 
 \;, \la{Seq} \\ 
 \lim_{\vc{r},\vc{r}'\to \vc{0}}
 \im \bigl[
 \nabla^{ }_{\vc{r}} \cdot \vc{G}(E';\vc{r},\vc{r}')
 \bigr]
 & \equiv & \rho^{ }_p(E') 
 \;, \la{rho}
\ea
where $\rho^{ }_p(E')$ is a spectral function.
Following refs.~\cite{4quark_lattice,threshold} 
and carrying out the integral over the 
center-of-mass momentum $\vc{k}$ (cf.\ \eq\nr{com}), we then get
\ba
 \langle \rmO^{ }_p \rangle & = &  
 \biggl( \frac{M^{ }_\rmi{kin} T}{\pi} \biggr)^{3/2}
 e^{- 2 M^{ }_\rmi{rest} / T}
 \int_{-\Lambda}^{\infty}
 \! \frac{{\rm d}E'}{\pi} \, e^{-E'/T} \rho^{ }_p(E')
 \;, \la{Op_spec} \\
 \bar{S}^{ }_p & = &  
 \frac{2 M^{ }_\rmi{kin}}{3 N T}
 \biggl(\frac{4\pi}{M^{ }_\rmi{kin} T} \biggr)^{3/2}
 \int_{-\Lambda}^{\infty}
 \! \frac{{\rm d}E'}{\pi} \, e^{-E'/T} \rho^{ }_p(E')
 \;.  \la{Sp_spec}
\ea
Here $\alpha^2 M^{ }_\rmi{kin} \ll \Lambda \ll M^{ }_\rmi{rest}$ 
is a cutoff restricting the average to the non-relativistic regime. 
As our masses $M^{ }_\rmi{rest}$, $M^{ }_\rmi{kin}$
already include thermal corrections, 
$
 n^{ }_\rmi{eq} = (n^{ }_\rmi{eq})^{ }_\rmi{tree}
$
within our approximation, so that \eq\nr{Sp_spec} is obtained
by dividing \eq\nr{Op_spec} through \nr{Op_tree}.

Let us crosscheck that \eqs\nr{Seq}--\nr{Sp_spec} are correct
at tree-level. Setting $V^{ }_\T(r)\to 0$ and 
$\Gamma^{ }_\T(r) \to 0^+$, \eq\nr{Seq}
is easily solved in momentum space, yielding ($p \equiv |\vc{p}|$)
\be
 \rho^{ }_\rmi{$p$,tree}(E') \; \equiv \;  \frac{ N }{M_\rmi{kin}^2}
 \int_{\vc{p}} {p}^2 
 \pi\, \delta \Bigl( E' 
 - \frac{p^2}{M^{ }_\rmi{kin}} \Bigr)
 \; = \; 
 \frac{N M_\rmi{kin}^{1/2} 
 \theta(E') (E')^{3/2}
 }{4\pi} 
 \;. \la{rhofree}
\ee
Inserting into \eq\nr{Sp_spec} and carrying out the integral over
$E'$ indeed gives unity. 

Another limit in which $\rho^{ }_p(E')$ can be determined 
is a Coulombic potential,  
namely $V^{ }_\T(r) \to  - \alpha/r$ and $\Gamma^{ }_\T(r) \to 0^+$. 
Parametrizing $E' = M^{ }_\rmi{kin} v^2$, 
the above-threshold solution reads
\be
 \rho^{ }_p(E') = \rho^{ }_\rmi{$p$,tree}(E') \, S^{ }_p(v)
 \;, \la{rho_pert_Sommer}
\ee
where $S^{ }_p$ is a vacuum $p$-wave Sommerfeld factor, given by
(cf.,\ e.g.,\ refs.~\cite{iengo,cassel})
\be
 S^{ }_p(v) = S^{ }_s(v) \, 
 \biggl( 1 + \frac{\alpha^2}{4 v^2}\biggr)
 \;, \quad
 S^{ }_s(v) \; \equiv \; 
 \frac{\pi\alpha/v}{1-e^{-\pi\alpha/v}}
 \;. \la{Sommer_Coulomb}
\ee 
A large enhancement is observed for $v \ll \alpha$, in particular
$\lim_{E'\to 0} \rho^{ }_p(E') = N\alpha^3 M_\rmi{kin}^2/16$. This 
enhancement originates from an overlap with an $s$-wave radial
function (this is explained in footnote~\ref{fn:s}), 
and gives the dominant
above-threshold contribution to $\bar{S}^{ }_p$ if 
$T \lsim \alpha^2 M^{ }_\rmi{kin}$.

\begin{figure}[t]

\hspace*{-0.1cm}
\centerline{%
 \epsfysize=7.6cm\epsfbox{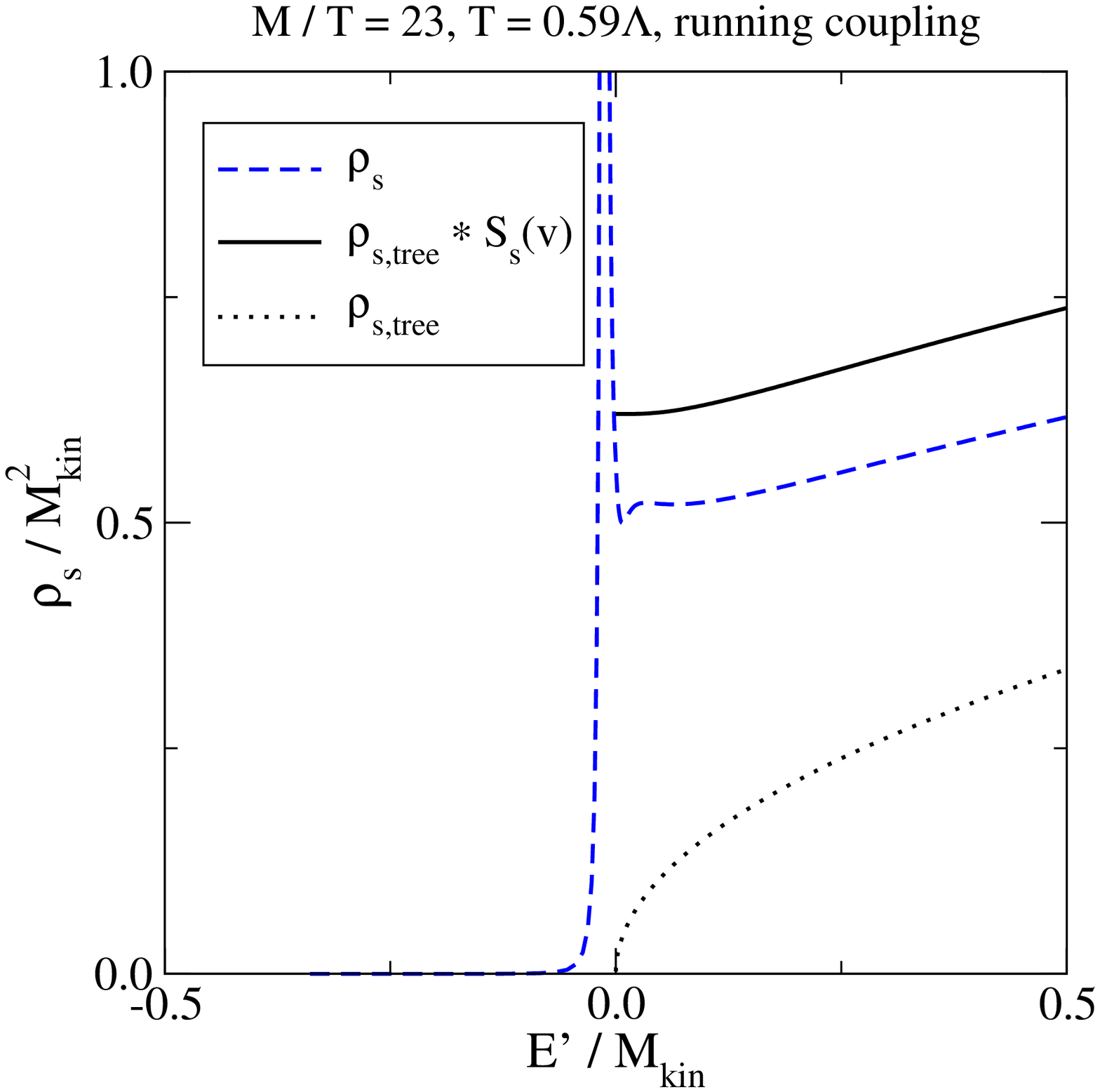}
 \hspace{0.1cm}
 \epsfysize=7.6cm\epsfbox{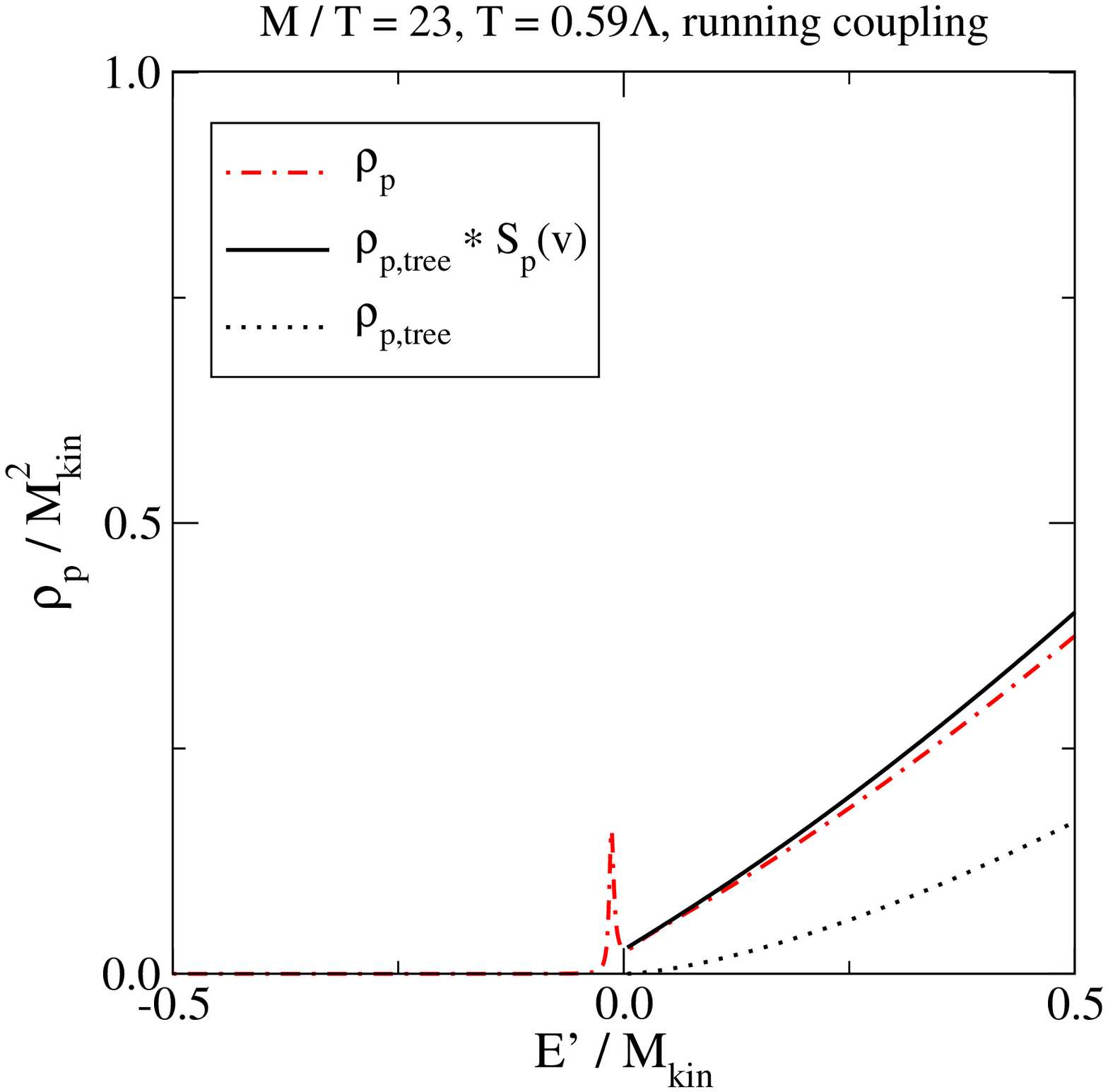}
}

\caption[a]{\small
 Left: 
 the perturbative $s$-wave spectral function from \eq\nr{peskin_rep}.
 Above the threshold, the result is compared with the prediction from
 a Coulombic Sommerfeld factor, cf.\ \eq\nr{Sommer_Coulomb}, 
 where $E' \equiv M^{ }_\rmi{kin} v^2$. 
 The correct result is below the
 prediction of the Sommerfeld factor, because the potential
 gets Debye-screened and because the running of the coupling reduces
 the coefficient of the attraction at short distances. 
 Right: 
 the same for the $p$-wave. 
}

\la{fig:rho_pert}
\end{figure}

A general numerical method
to find the solution of the $s$-wave analogues of 
\eqs\nr{Seq} and \nr{rho}
was presented in ref.~\cite{original}, 
and an implementation for the $p$-wave was worked 
out in ref.~\cite{peskin}. The solutions can be written as\footnote{%
 Let us elaborate on the origin of the two parts in $\rho^{ }_p$. 
 In terms of eigenstates 
 of the operator in \eq\nr{Seq}, the $p$-wave solution contains
 $\nabla \psi (\vc{0})$. In spherical coordinates, 
 writing $\psi = R^{ }_{nl}(r) Y^{ }_{lm}(\Omega)$, 
 we thus need $R_{nl}'(0)$. In a Coulomb potential, $R^{ }_{n0}$ has
 a linear term at small $r$,
 $R^{ }_{n0}(r) = c^{ }_0 + c^{ }_1 r + ...$, which leads to 
 an $s$-wave contribution to $\nabla \psi (\vc{0})$, 
 denoted in \eq\nr{peskin_rep} by 
 $u^{ }_0 \sim r R^{ }_{n0}$. 
 This is responsible for the dominant term $\sim \alpha^2/(4v^2)$ 
 in \eq\nr{Sommer_Coulomb}. 
 The second term in $\rho^{ }_p$ of \eq\nr{peskin_rep} is the ``genuine''
 $p$-wave contribution, originating from the short-distance asymptotics 
 $R^{ }_{n1}(r) = d^{ }_0 r + ...$. \la{overlap} \la{fn:s}
 } 
\be
 \frac{\rho^{ }_s(E')}{M_\rmi{kin}^2} 
 \; = \; 
 \frac{N \alpha}{4\pi}
 \int_0^\infty \! {\rm d}\rho \, 
 \im \biggl( 
   \frac{1}{u_0^2}
 \biggr)
 \;, 
 \quad
 \frac{\rho^{ }_p(E')}{M_\rmi{kin}^2} 
 \; = \; 
 \frac{N \alpha^3}{16\pi}
 \int_0^\infty \! {\rm d}\rho \, 
 \im \biggl( 
   \frac{1}{u_0^2}
 + \frac{36}{u_1^2}
 \biggr)^{ }_{ }
 \;, \la{peskin_rep}
\ee
where $\alpha \equiv g_s^2 \CF/(4\pi)$, 
$\rho \equiv r \alpha M^{ }_\rmi{kin}$, and $u^{ }_\ell$ 
is a regular solution of the homogeneous equation
\be
 \biggl\{
 \frac{\partial^2}{\partial \rho^2} - \frac{\ell(\ell+1)}{\rho^2}
 + \frac{E'+ i \Gamma^{ }_\T(r) - V^{ }_\T(r)}{M^{ }_\rmi{kin} \alpha^2} 
 \biggr\} \, u^{ }_\ell(\rho) = 0 
 \;, \la{u_ell}
\ee
assumed normalized as 
$
 u^{ }_\ell = \rho^{\ell + 1} + ...
$
at short distances. 
It is sufficient to solve the equation
up to some finite $\rho \gg 1$ and attach this 
to the known asymptotics.\footnote{%
 For $\rho \gg 1$, $V^{ }_T$ vanishes and 
 $\Gamma^{ }_\T$ goes over to a constant, whereby the 
 equation satisfied by the $p$-wave wave function reads 
 $
  (\partial_\rho^2 - \frac{2}{\rho^2}+ \hat E' + i \hat \Gamma) u^{ }_1 = 0
 $, 
 where $\hat E' \equiv E'/ (M^{ }_\rmii{kin} \alpha^2)$
 and $\hat \Gamma \equiv \Gamma^{ }_\T (\infty)/(M^{ }_\rmii{kin} \alpha^2)$. 
 Let us denote $k \equiv \sqrt{\hat E' + i \hat \Gamma}$. 
 Then the general solution reads
 $
 u^{ }_1 = C [ \sin(k \rho + \delta) / (k \rho) -
  \cos(k \rho + \delta) ]  
 $, 
 where $C,\delta \in \mathbbm{C}$.  
 The function $1/u_1^2$ is integrable, and subsequently 
 $C \sin(k \rho + \delta)$ and 
 $C \cos(k \rho + \delta)$ can be 
 expressed in terms of $u^{ }_1(\rho)$ and 
 $u'_1(\rho)$. 
 We thus obtain
 \be
  \int_{\rho^{ }_0}^\infty \! {\rm d}\rho \, 
  \im \biggl( \frac{1}{u_1^2} \biggr) 
  = 
  \im \Biggl\{ 
  \frac{1}{u^{ }_1(\rho^{ }_0)
  \bigl[ 
   u'_1(\rho^{ }_0)  
   + u^{ }_1(\rho^{ }_0)  
   \frac{1 - i k^3 \rho_0^3}{\rho^{ }_0(1+k^2 \rho_0^2)}
  \bigr]}
  \Biggr\} 
  \;. 
 \ee
 Setting $\rho^{ }_0 \to \epsilon\equiv 0^+$ and recalling 
 $ u^{ }_1(\epsilon) \approx \epsilon^{2}$ yields $\re(k^3/9)$,
 which reproduces \eq\nr{rhofree} from \eq\nr{peskin_rep}.
 The part
 $
  \int_{\rho^{ }_0}^\infty \! {\rm d}\rho \, 
  \im \bigl( {1}/{u_0^2} \bigr) 
 $
 of \eq\nr{peskin_rep}
 yields $\re(k)$ when $\rho^{ }_0 \to \epsilon$, 
 which amounts to 
 $ 
   \sim \alpha^2 \theta(E')(E')^{1/2}
 $.  
}

As far as the potential goes, at large separations we make use of 
a Hard Thermal Loop resummed thermal expression which includes the 
effects of Debye screening and Landau damping~\cite{imV,bbr,jacopo},
\be
 V^{ }_\T(r)  =  - \frac{g_s^2 \CF^{ } \exp(-\mD^{ } r)}{4\pi r}
 \;, \quad 
 \Gamma^{ }_\T(r)  =  \frac{g_s^2 \CF^{ } T}{2\pi}
 \int_0^\infty \! \frac{{\rm d} z \, z}{(z^2 +1)^2}
 \biggl[
   1 - \frac{\sin(z \mD^{ }r)}{z \mD^{ }r} 
 \biggr]
 \;, \la{imV}
\ee
where $\mD^{ } \sim g^{ }_sT $ is a Debye mass.
For numerical estimates we insert 
2-loop values of $\mD^{ }$ and 
$g_s^2$ from ref.~\cite{gE2}
(the 3-loop level has been reached only for $\mD^{ }$~\cite{mE2}).  
The real part of the potential is interpolated into a vacuum
expression at short separations~\cite{pot1,pot4}, 
as discussed in ref.~\cite{GPtau}. 
In order to account for the proper kinematics of real processes
in a regime beyond which the derivation is formally valid, we 
also follow the arguments presented in ref.~\cite{idm} and
multiply the imaginary part of the potential by the Boltzmann 
factor $e^{-|E'|/T}$ for $E' < 0$. Corresponding numerical
solutions of the spectral functions $\rho^{ }_s$ and $\rho^{ }_p$
are shown in \fig\ref{fig:rho_pert}.

%
\section{Lattice framework}
\la{se:latt}

On the lattice the double ratio in \eq\nr{Si_def} is replaced through 
\be
 \bar{S}^{ }_{p} \; \equiv \; 
 \frac{P^{ }_{p} / P^\rmi{cold}_{p} }{ ( P^{ }_1 / P^\rmi{cold}_1 )^2 }
 \;, \la{Sp_latt}
\ee
where $P^{ }_1$ and $P^{ }_p$ are expectation values to be specified
presently (cf.\ \eqs\nr{P1} and \nr{Pp}). The superscript ``cold'' 
indicates a measurement with all link matrices set to 
unity; this is an implementation of the ``tree-level'' prescription
of perturbation theory. The division by the respective cold measurement
implies that $\bar{S}^{ }_p$ deviates from unity only through the effect
of gauge interactions. The normalization by $P^{2}_1$ furthermore 
implies that modifications of the rest mass by 
gauge interactions are cancelled, an effect
which is linearly divergent in lattice regularization and strongly
influences $n^{ }_\rmi{eq}$ (cf.\ \eq\nr{neq}). 

For a lattice measurement, we choose a simple first-order discretization
of the covariant derivatives in \eq\nr{O_p}. We denote by $U^{ }_i$ a
link in the $i^\rmi{th}$ direction with origin at $\vc{0}$, 
by $\vc{i} \equiv a^{ }_s \vc{e}_i$ a displacement 
in the $i^\rmi{th}$ direction by a lattice spacing $a^{ }_s$, 
and by $G^\theta_{ }$, $G^\chi_{ }$ the propagators
\ba
 G^\theta_{\alpha\gamma;kl}(\tau_2,\vc{x};\tau_1,\vc{y}) & \equiv & 
 \bigl\langle 
 \theta^{ }_{\alpha k}( \tau_2,\vc{x} )\, 
 \theta^\dagger_{\gamma l} (\tau_1,\vc{y} ) 
 \bigr\rangle
 \;, \la{G_theta} \\
 G^\chi_{\alpha\gamma;kl}(\tau_2,\vc{x};\tau_1,\vc{y}) & \equiv & 
 \bigl\langle 
 \chi^{ }_{\alpha k}( \tau_2,\vc{x} )\, 
 \chi^\dagger_{\gamma l} (\tau_1,\vc{y} ) 
 \bigr\rangle
 \;, \la{G_chi}
\ea
where $\alpha,\gamma \in \{1,...,\Nc^{ }\}$ are colour indices
and $k,l\in\{1,2\}$ are spin indices. Given that $\chi$ represents
an antiparticle to $\theta$, the two propagators are related by
\be
 G^\chi_{ }(\tau_2,\vc{x};\tau_1,\vc{y})
 = - 
 \bigl[ 
 G^\theta_{ }(\tau_1,\vc{y};\tau_2,\vc{x})
 \bigr]^\dagger
 \;. 
\ee
Because non-relativistic particles move in the positive 
time direction only, a non-zero contraction
may necessitate propagating
across the imaginary time interval, whose extent is $\beta \equiv 1/T$. 
For taking derivatives of a propagator
with respect to the position of a sink or source we introduce
a shorthand notation,  
\be
 D^{ }_i G^\theta_{\alpha\gamma;kl} 
 \; \equiv \; 
 \bigl\langle 
 (D^{ }_i\theta)^{ }_{\alpha k}( \beta,\vc{x} )\, 
 \theta^\dagger_{\gamma l} (0,\vc{x} ) 
 \bigr\rangle
 \;, \quad 
 G^\theta_{\alpha\gamma;kl;i} 
 \; \equiv \; 
 \bigl\langle 
 \theta^{ }_{\alpha k}( \beta,\vc{x} )\, 
 (D^{ }_i \theta) ^\dagger_{\gamma l} (0,\vc{x} ) 
 \bigr\rangle
 \;. \la{Gtheta_i}
\ee

With these propagators, the lattice analogue of $n^{ }_\rmi{eq}$ 
reads~\cite{4quark_lattice}
\be
 \bigl( n^{ }_\rmi{eq} \bigr)^{ }_\rmi{latt}
 \; = \; 
 2 \re \tr \bigl\langle 
   G^\theta(\beta,\vc{0};0,\vc{0}) \bigr\rangle 
 \;. \la{neq_latt}
\ee
Given that overall normalization cancels out in \eq\nr{Sp_latt}, we 
in practice define $P^{ }_1$ by dividing 
$( n^{ }_\rmi{eq} )^{ }_\rmi{latt}$ by the number
of degrees of freedom, {\it viz.}\ $2N$, i.e.\ 
\ba
 P^{ }_1 & \equiv & \frac{1}{N} \re\, \bigl\langle 
   G^\theta_{\alpha\alpha;ii}(\beta,\vc{0};0,\vc{0})  \bigr\rangle 
 \;. \la{P1}
\ea
For the operator in \eq\nr{O_p}, Wick contractions yield
\be
 \langle \rmO^{ }_p \rangle  =  
 \frac{1}{2 M_\rmi{kin}^2}
 \sum_{i=1}^{3} \re\tr 
 \bigl\langle
   D^{ }_i G^\theta_{;i} G^{\theta\dagger}_{ } 
   \, - \, 
   D^{ }_i G^\theta_{ } G^{\theta\dagger}_{;i} 
 \bigr\rangle
 \;.
\ee
Replacing covariant derivatives by discrete lattice derivatives, and
choosing again a convenient normalization, whose effects cancel out
in \eq\nr{Sp_latt}, we are led to define
\ba
 P^{ }_p 
 & \equiv & 
 \frac{1}{2N}
 \sum_{i=1}^{3} \re\tr\, 
 \Bigl\langle
   G^\theta_{ }(\beta,\vc{i};0,\vc{i})\,
   U^{\dagger}_i\,
   G^{\theta\dagger}_{ }(\beta,\vc{0};0,\vc{0})\,
   U^{ }_i
 \hphantom{\;.}
 \;\quad\; \NRa
 \nn 
 & & \hspace*{2.15cm} - \, 
   G^\theta_{ }(\beta,\vc{i};0,\vc{0})\,
   U^{ }_i\,
   G^{\theta\dagger}_{ }(\beta,\vc{0};0,\vc{i})\,
   U^{ }_i
 \Bigr\rangle
 \;.  
 \;\quad \NRb  \la{Pp}
\ea
The diagrams illustrate the topology of the contractions. 

The lattice framework and the gauge ensemble 
are as in ref.~\cite{4quark_lattice}. The light
sector consists of SU(3) gauge theory and $\Nf = 2+1$ flavours of vectorlike
fermions transforming in the fundamental representation. The parameters
of the action were tuned in refs.~\cite{lat0a,lat0b}. Denoting by 
$\Lambda$ 
a scale parameter~(cf.~ref.~\cite{Lambda} for a review), 
the lightest pseudoscalar mesons have masses 
$1.2 \Lambda$ 
and $1.5 \Lambda$, 
respectively, the latter for
the mesons involving one quark of the third flavour. 
The lattice is anisotropic, with 
$a^{ }_s/a^{ }_\tau \approx 3.5$, where 
the spatial lattice spacing is 
$a^{ }_s \approx 0.21 \Lambda^{-1}$. 
The spatial extent of the box is $L = 24 a^{ }_s$. 
The system is put at a finite temperature by tuning $N_\tau$, 
i.e.\ the number of temporal lattice sites, so that 
$T = \frac{1}{N_\tau a_\tau}$.
The system has a (pseudo)critical temperature at 
$\Tc \approx 0.54 \Lambda$~\cite{lat1a}. 
Thermal properties of the system were studied
in ref.~\cite{lat1b}. 
We vary $T = (0.95 ... 1.9)\Tc^{ }$ and, 
setting 
$ M^{ }_\rmi{kin} = 14 \Lambda $, 
can hence access 
values $M^{ }_\rmi{kin}/T\sim 14 ... 28$, a reasonable range in view of 
dark matter freeze-out computations. 

%
\section{Numerical results and their uncertainties}
\la{se:num}

\begin{figure}[t]

\hspace*{-0.1cm}
\centerline{%
 \epsfxsize=7.6cm\epsfbox{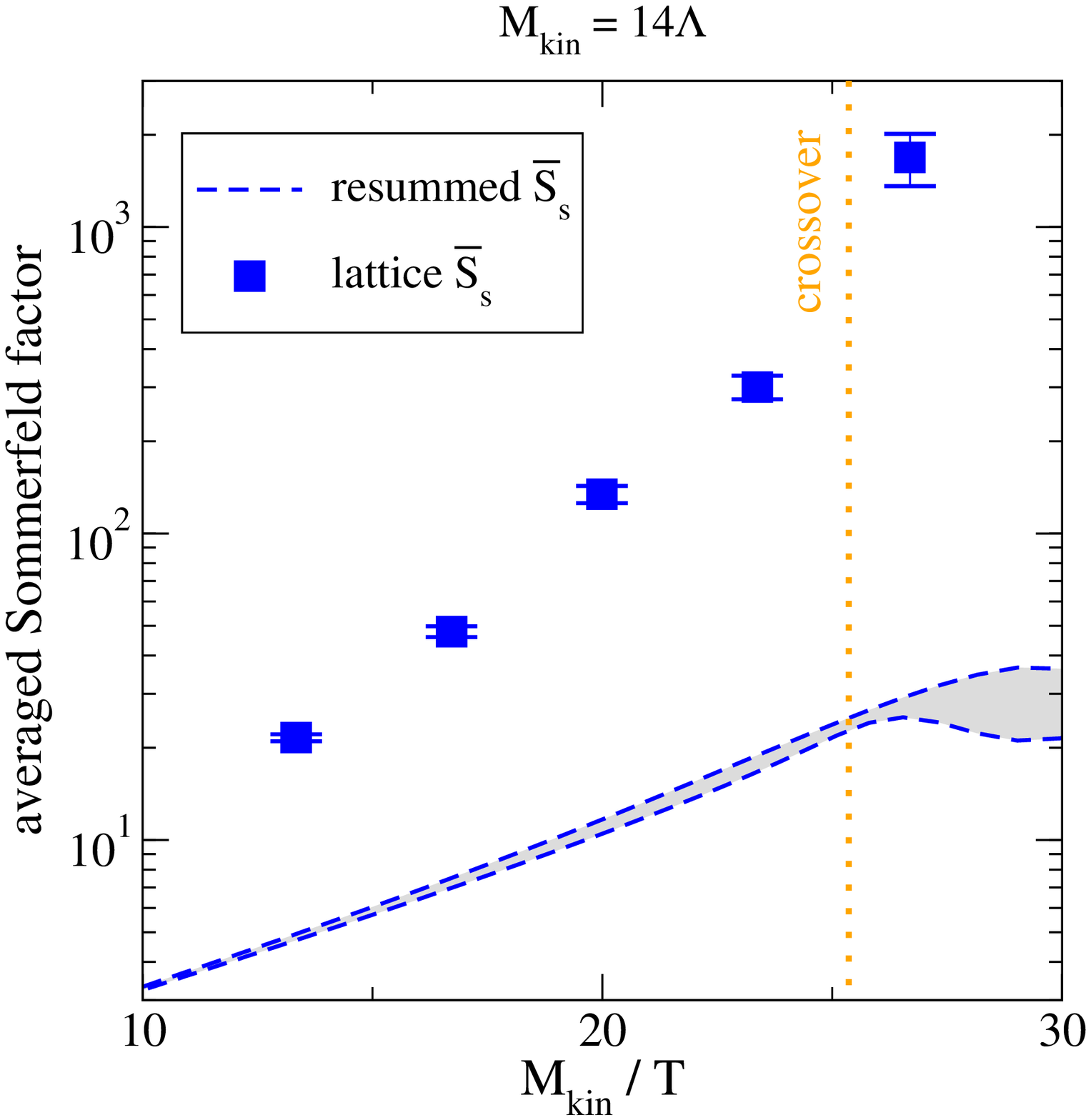}
 \hspace{0.1cm}
 \epsfxsize=7.6cm\epsfbox{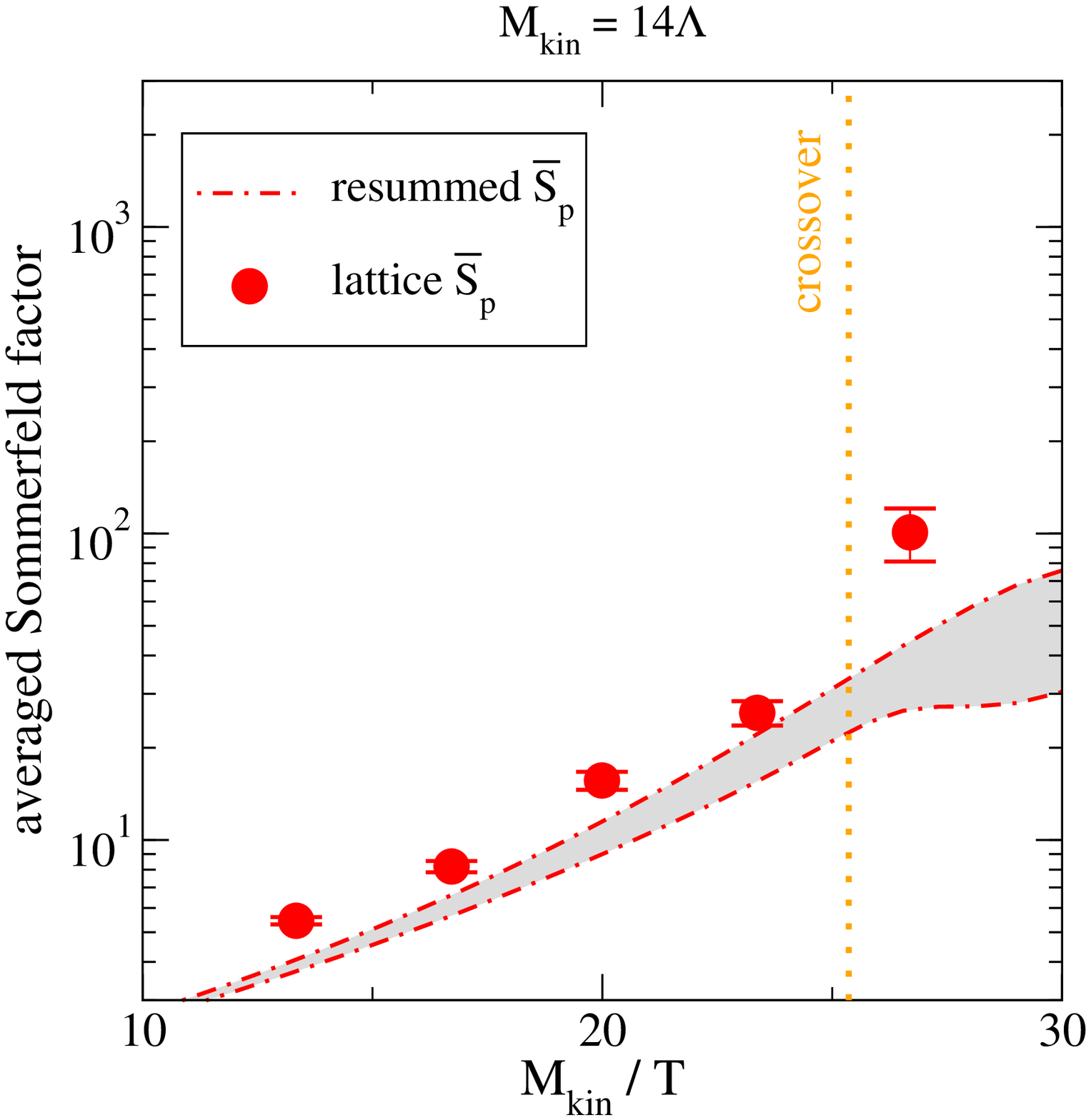}%
}

\caption[a]{\small
 Left: 
 thermally averaged Sommerfeld factors for the $s$-wave~\cite{4quark_lattice}. 
 Right: 
 the same for the $p$-wave. 
 Grey bands represent scale uncertainties of resummed 
 perturbation theory, cf.\ \se\ref{se:pert}, whereas the error bars
 show statistical errors of lattice simulations, cf.\ \se\ref{se:latt}. 
 As discussed in \se\ref{se:num}, the true systematic uncertainties
 are much larger on both sides. The vertical dashed line
 indicates the crossover at which the systems goes into a confined phase.  
}

\la{fig:sommer}
\end{figure}

Perturbative results 
for thermally averaged Sommerfeld factors
from \se\ref{se:pert} and lattice results from \se\ref{se:latt}  
are compared with each other in \fig\ref{fig:sommer}
(the errors shown for the lattice results are statistical only). 
For the $p$-wave, shown in  
\fig\ref{fig:sommer}(right), we find surprisingly good qualitative 
agreement, indicating an enhancement factor $\sim 100$ at the lowest
temperature. We note that the system is in a confined phase for 
$M/T \gsim 26$. 

For the $s$-wave, shown in \fig\ref{fig:sommer}(left), 
the discrepancy between the perturbative and lattice 
results is rather substantial.\footnote{%
 In ref.~\cite{4quark_lattice}, 
 the perturbative values were noticeably 
 larger, and the agreement looked better. There are two reasons for this:  
 in ref.~\cite{4quark_lattice} we used the larger 1-loop thermal coupling, 
 and most importantly the Salpeter correction 
 (thermal shift of the threshold location to smaller energies) 
 was included in the Sommerfeld factor on the perturbative side.
 The latter has now be 
 excluded from the definition of the Sommerfeld factor through
 \eq\nr{Si_def} on both the perturbative and lattice side,
 so we believe the comparison to be fairer. 
} 
In fact, naively $\bar{S}^{ }_p > \bar{S}^{ }_s$
(cf.\ \eq\nr{Sommer_Coulomb}), whereas on the lattice 
$\bar{S}^{ }_s$ clearly exceeds $\bar{S}^{ }_p$. 
In this context we note that physically, 
the thermally averaged Sommerfeld factors are sensitive both
to energy levels and to the corresponding ``overlaps'', 
or wave functions at origin
($|\psi(\vc{0})|^2$, $|\nabla\psi(\vc{0})|^2$). 
For another observable in a similar temperature range, 
it has been found that while for energy levels there is 
fair agreement, lattice and perturbative
overlaps show substantial discrepancies
(cf.\ fig.~6 in ref.~\cite{screening}).

Let us discuss possible reasons for the discrepancy. Starting with 
the perturbative side, we are quite close to the confined phase and
correspondingly our effective coupling is large, varying in the range 
$\alpha^{ }_s \simeq 0.3 ... 0.6$
for $M^{ }_\rmi{kin}/T \simeq 10 ... 30$. 
The grey bands in \fig\ref{fig:sommer}
originate from the variation of a thermal $\alpha^{ }_s$~\cite{gE2} 
as we change 
the renormalization scale within a factor $\fr12 ... 2$.
In the $s$-wave case, the corresponding error band looks quite narrow. 
The reason is that in this parameter range the value of $\bar{S}^{ }_s$ 
is influenced by above-threshold scattering states, i.e.\ tree-level 
processes, which are insensitive to $\alpha^{ }_s$. 
If we artificially increase $\alpha^{ }_s$ by a factor two, into  
the range $0.6 ... 1.2$, then $\bar{S}^{ }_s$ increases by 
a factor $3 ... 20$, improving the agreement, 
however $\bar{S}^{ }_p$ increases simultaneously by a factor $4 ... 70$, 
spoiling the agreement on that side.  
In principle a possible way to reduce these uncertainties would be 
a systematic higher-order computation, however it represents  
a daunting task, including the need for a careful 
power counting concerning which resummations are necessary 
in the various temperature and mass ranges of interest. 

On the lattice side, no infinite-volume
or continuum extrapolation was carried out.
A box of a finite size influences  
the spectrum of scattering states, and given that scattering 
states contribute to the pair annihilation process, this might
imply the presence of finite-volume effects. 
If the system has tightly bound states, whose Bohr radius is not
much larger than the lattice spacing, there may also be large 
discretization effects.   
In order to check whether the lattice results are plagued by
finite-volume or discretization artifacts, additional sets
of simulations are needed, requiring a major computational effort
beyond our resources. 

%
\section{Conclusions}
\la{se:concl}

Building upon a framework developed in ref.~\cite{4quark_lattice}, 
we have estimated the thermally averaged $p$-wave Sommerfeld
factor associated with a particular annihilation channel
(cf.\ \eq\nr{O_p}), both through a resummed perturbative 
(cf.\ \se\ref{se:pert}) and through a lattice
computation (cf.\ \se\ref{se:latt}). 
Both methods suggest that large enhancement factors
$\sim 100$ are possible (cf.\ \fig\ref{fig:sommer}). 

Within naive perturbation theory, $\bar{S}^{ }_p > \bar{S}^{ }_s$
(cf.\ \eq\nr{Sommer_Coulomb}), 
but on the lattice we find $\bar{S}^{ }_s > \bar{S}^{ }_p$
(cf.\ \fig\ref{fig:sommer}). 
We may speculate that the large 
non-perturbative increase of $\bar{S}^{ }_s$ is due to 
more prominent bound-state effects in the $s$-wave, however
systematic uncertainties may also play a role (cf.\ \se\ref{se:num}), 
an effect which can hopefully be clarified through future work.

In cosmological applications, with $M \gsim 1$~TeV, we normally find 
ourselves in the regime $T \gg \Lambda$, which implies that $\alpha$ is 
smaller than in our study. However, as indicated by \eq\nr{Sommer_Coulomb}, 
the magnitude of the averaged Sommerfeld factors depends on the ratio 
$\sim \langle \alpha/v \rangle \sim \sqrt{\alpha^2 M^{ }_{\rm kin}/T}$.
Therefore large averaged Sommerfeld factors are found at least in the regime 
$M^{ }_{\rm kin}/T \gg 100$, relevant for late-time pair annihilations.
Because of the smaller $\alpha$, higher order corrections should be 
smaller than in our study. The fact that we find qualitative resemblances 
even in our \fig\ref{fig:sommer}, then suggests that resummed perturbative 
estimates should be conservative in that case. 
For $M^{ }_\rmi{kin}/T \in (10,1000)$, resummed perturbative
values of $\bar{S}^{ }_s$ from ref.~\cite{stop} can be found
on the web site {\tt http://www.laine.itp.unibe.ch/sommerfeld}, and 
we have now added corresponding results for $\bar{S}^{ }_p$ there. 

%
\section*{Acknowledgements}

We thank the FASTSUM collaboration for providing the gauge
configurations used in our measurements. M.L.\ thanks Maxim Khlopov
for interesting correspondence. 
S.K.\ was
supported by the National Research Foundation of Korea under grant
No.\ 2018R1A2A2A05018231 funded by the Korean government (MEST) and in
part by NRF-2008-000458. M.L.\ was supported 
by the Swiss National Science Foundation
(SNF) under grant 200020-168988.

\small{
%

}


\begin{thebibliography}{99}

\bibitem{lsb}
  L.S.~Brown and R.F.~Sawyer,
  {\it Nuclear reaction rates in a plasma,}
  Rev.\ Mod.\ Phys.\  {69} (1997) 411
  [astro-ph/9610256].

\bibitem{bodwin}
  G.T.~Bodwin, E.~Braaten and G.P.~Lepage,
  {\it Rigorous QCD analysis of inclusive annihilation and 
  production of heavy quarkonium,}
  Phys.\ Rev.\ D {51} (1995) 1125; 
  {\it ibid.} {55} (1997) 5853 (E)
  [hep-ph/9407339].

\bibitem{asommerfeld} 
  A.~Sommerfeld, 
  {\em \"Uber die Beugung und Bremsung der Elektronen}, 
  Ann.\ Phys.\ (Leipzig) {403} (1931) 257.

\bibitem{landau3} 
  L.D.~Landau and E.M.~Lifshitz, 
  {\it Quantum Mechanics, Non-Relativistic Theory,} Third Edition, \S136
  (Butterworth-Heinemann, Oxford).

\bibitem{gamow}
  G.~Gamow, 
  {\it Zur Quantentheorie des Atomkernes,}
  Z.\ Physik 51 (1928) 204.

\bibitem{sakharov} 
  A.D.~Sakharov,
  {\it Interaction of an Electron and Positron in Pair Production,}
  Zh.\ Eksp.\ Teor.\ Fiz.\  {18} (1948) 631
  [Sov.\ Phys.\ Usp.\  {34} (1991) 375]

\bibitem{hisano}
  J.~Hisano, S.~Matsumoto, M.M.~Nojiri and O.~Saito,
  {\it Non-perturbative effect on dark matter annihilation and gamma
  ray signature from galactic center,}
  Phys.\ Rev.\ D {71} (2005) 063528
  [hep-ph/0412403].

\bibitem{sfeldx}
  M.~Cirelli, A.~Strumia and M.~Tamburini,
  {\it Cosmology and Astrophysics of Minimal Dark Matter,}
  Nucl.\ Phys.\ B {787} (2007) 152
  [0706.4071].

\bibitem{feng}
  J.L.~Feng, M.~Kaplinghat and H.-B.~Yu,
  {\em Sommerfeld Enhancements for Thermal Relic Dark Matter,}
  Phys.\ Rev.\ D {82} (2010) 083525
  [1005.4678].

\bibitem{mb} 
  M.~Beneke, C.~Hellmann and P.~Ruiz-Femen\'ia,
  {\it Heavy neutralino relic abundance with Sommerfeld enhancements --
  a study of pMSSM scenarios,}
  JHEP {03} (2015) 162
  [1411.6930].

\bibitem{etc}
  S.~El Hedri, A.~Kaminska and M.~de Vries,
  {\it A Sommerfeld Toolbox for Colored Dark Sectors,}
  Eur.\ Phys.\ J.\ C {77} (2017) 622
  [1612.02825].

\bibitem{old32}
  W.~Detmold, M.~McCullough and A.~Pochinsky,
  {\it Dark Nuclei I: Cosmology and Indirect Detection,}
  Phys.\ Rev.\ D {90} (2014) 115013
  [1406.2276].

\bibitem{old4}
  B.~von Harling and K.~Petraki,
  {\it Bound-state formation for thermal relic dark matter and unitarity,}
  JCAP {12} (2014) 033
  [1407.7874].

\bibitem{giv}
  M.~Garny, A.~Ibarra and S.~Vogl,
  {\it Signatures of Majorana dark matter with $t$-channel mediators,}
  Int.\ J.\ Mod.\ Phys.\ D {24} (2015) 1530019
  [1503.01500].

\bibitem{mg3}
  M.~Garny, J.~Heisig, M.~Hufnagel and B.~L\"ulf,
  {\it Top-philic dark matter within and beyond the WIMP paradigm,}
  Phys.\ Rev.\ D {97} (2018) 075002
  [1802.00814].

\bibitem{4quark_lattice}
  S.~Kim and M.~Laine,
  {\it Rapid thermal co-annihilation through bound states in QCD,}
  JHEP {07} (2016) 143
  [1602.08105].

\bibitem{threshold}
  S.~Kim and M.~Laine,
  {\it On thermal corrections to near-threshold annihilation,}
  JCAP {01} (2017) 013
  [1609.00474].

\bibitem{stop}
  S.~Biondini and M.~Laine,
  {\it Thermal dark matter co-annihilating with 
  a strongly interacting scalar,}
  JHEP {04} (2018) 072
  [1801.05821].

\bibitem{sb}
  S.~Biondini and S.~Vogl,
  {\it Coloured coannihilations: 
  Dark matter phenomenology meets non-relativistic EFTs,}
  JHEP {02} (2019) 016
  [1811.02581].

\bibitem{ll}
  S.P.~Liew and F.~Luo,
  {\it Effects of QCD bound states on dark matter relic abundance,}
  JHEP {02} (2017) 091
  [1611.08133].

\bibitem{mrss}
  A.~Mitridate, M.~Redi, J.~Smirnov and A.~Strumia,
  {\it Cosmological Implications of Dark Matter Bound States,}
  JCAP {05} (2017) 006
  [1702.01141].

\bibitem{klz}
  W.Y.~Keung, I.~Low and Y.~Zhang,
  {\it A Reappraisal on Dark Matter Co-annihilating with a Top/Bottom
  Partner,}
  Phys.\ Rev.\ D {96} (2017) 015008
  [1703.02977].

\bibitem{hp}
  J.~Harz and K.~Petraki,
  {\it Radiative bound-state formation in unbroken perturbative 
  non-Abelian theories and implications for dark matter,}
  JHEP {07} (2018) 096
  [1805.01200].

\bibitem{iengo} 
  R.~Iengo,
  {\it Sommerfeld enhancement: General results from field theory diagrams,}
  JHEP {05} (2009) 024
  [0902.0688].

\bibitem{cassel}
  S.~Cassel,
  {\it Sommerfeld factor for arbitrary partial wave processes,}
  J.\ Phys.\ G {37} (2010) 105009
  [0903.5307].

\bibitem{hg}
  H.~Goldberg,
  {\it Constraint on the Photino Mass from Cosmology,}
  Phys.\ Rev.\ Lett.\  {50} (1983) 1419;
  {\it ibid.} {103} (2009) 099905 (E). 

\bibitem{Zhao:2016xie}
  Y.~Zhao, X.J.~Bi, H.Y.~Jia, P.F.~Yin and F.R.~Zhu,
  {\it Constraint on the velocity dependent dark matter annihilation
  cross section from Fermi-LAT observations of dwarf galaxies,}
  Phys.\ Rev.\ D {93} (2016) 083513
  [1601.02181].

\bibitem{Choquette:2016xsw}
  J.~Choquette, J.M.~Cline and J.M.~Cornell,
  {\it p-wave Annihilating Dark Matter from a Decaying Predecessor
  and the Galactic Center Excess,}
  Phys.\ Rev.\ D {94} (2016) 015018
  [1604.01039].

\bibitem{An:2016kie}
  H.~An, M.B.~Wise and Y.~Zhang,
  {\it Strong CMB Constraint On P-Wave Annihilating Dark Matter,}
  Phys.\ Lett.\ B {773} (2017) 121
  [1606.02305].

\bibitem{clas1}
  B.W.~Lee and S.~Weinberg, 
  {\it Cosmological Lower Bound on Heavy Neutrino Masses}, 
  Phys.\ Rev.\ Lett.\  {39} (1977) 165.

\bibitem{clas2}
  J.~Bernstein, L.S.~Brown and G.~Feinberg,
  {\it The Cosmological Heavy Neutrino Problem Revisited,}
  Phys.\ Rev.\ D {32} (1985) 3261.

\bibitem{old1}
  K.~Griest and D.~Seckel,
  {\it Three exceptions in the calculation of relic abundances,}
  Phys.\ Rev.\ D {43} (1991) 3191.

\bibitem{chemical}
  D.~B\"odeker and M.~Laine,
  {\it Heavy quark chemical equilibration rate as a transport coefficient,}
  JHEP {07} (2012) 130
  [1205.4987].

\bibitem{binder}
  T.~Binder, L.~Covi and K.~Mukaida,
  {\it Dark Matter Sommerfeld-enhanced annihilation and Bound-state 
  decay at finite temperature,}
  Phys.\ Rev.\ D {98} (2018) 115023
  [1808.06472].

\bibitem{chesler}
  P.M.~Chesler, A.~Gynther and A.~Vuorinen,
  {\it On the dispersion of fundamental particles in 
  QCD and $\mathcal{N}=4$ Super Yang-Mills theory,}
  JHEP {09} (2009) 003
  [0906.3052].

\bibitem{original}
  M.J.~Strassler and M.E.~Peskin,
  {\it Threshold production of heavy top quarks: QCD and the Higgs boson,}
  Phys.\ Rev.\ D {43} (1991) 1500.

\bibitem{peskin}
  Y.~Burnier, M.~Laine and M.~Veps\"al\"ainen,
  {\it Heavy quarkonium in any channel in resummed hot QCD,}
  JHEP {01} (2008) 043
  [0711.1743].

\bibitem{imV}
  M.~Laine, O.~Philipsen, P.~Romatschke and M.~Tassler,
  {\it Real-time static potential in hot QCD,}
  JHEP {03} (2007) 054
  [hep-ph/0611300].

\bibitem{bbr}
  A.~Beraudo, J.-P.~Blaizot and C.~Ratti,
  {\it Real and imaginary-time $Q\overline{Q}$
  correlators in a thermal medium,}
  Nucl.\ Phys.\ A {806} (2008) 312
  [0712.4394].

\bibitem{jacopo}
  N.~Brambilla, J.~Ghiglieri, A.~Vairo and P.~Petreczky,
  {\it Static quark-antiquark pairs at finite temperature,}
  Phys.\ Rev.\ D {78} (2008) 014017
  [0804.0993].

\bibitem{gE2}
  M.~Laine and Y.~Schr\"oder,
  {\it Two-loop QCD gauge coupling at high temperatures,}
  JHEP {03} (2005) 067
  [hep-ph/0503061].

\bibitem{mE2}
  I.~Ghi\c{s}oiu, J.~M\"oller and Y.~Schr\"oder,
  {\it Debye screening mass of hot Yang-Mills theory to three-loop order,}
  JHEP {11} (2015) 121
  [1509.08727].

\bibitem{pot1}
  Y.~Schr\"oder,
  {\it The Static potential in QCD to two loops,}
  Phys.\ Lett.\ B {447} (1999) 321
  [hep-ph/9812205].

\bibitem{pot4}
  R.N.~Lee, A.V.~Smirnov, V.A.~Smirnov and M.~Steinhauser,
  {\it Analytic three-loop static potential,}
  Phys.\ Rev.\ D {94} (2016) 054029
  [1608.02603].

\bibitem{GPtau}
  Y.~Burnier, H.-T.~Ding, O.~Kaczmarek, A.-L.~Kruse, 
  M.~Laine, H.~Ohno and H.~Sandmeyer,
  {\it Thermal quarkonium physics in the pseudoscalar channel,}
  JHEP {11} (2017) 206
  [1709.07612].

\bibitem{idm}
  S.~Biondini and M.~Laine,
  {\it Re-derived overclosure bound for the inert doublet model,}
  JHEP {08} (2017) 047
  [1706.01894].

\bibitem{lat0a}
  R.G.~Edwards, B.~Joo and H.W.~Lin,
  {\it Tuning for Three-flavors of Anisotropic Clover Fermions
  with Stout-link Smearing,}
  Phys.\ Rev.\ D {78} (2008) 054501
  [0803.3960].

\bibitem{lat0b}
  H.W.~Lin {\it et al.} [Hadron Spectrum Collaboration],
  {\it First results from 2+1 dynamical quark flavors on an anisotropic
  lattice: Light-hadron spectroscopy and setting the strange-quark mass,}
  Phys.\ Rev.\ D {79} (2009) 034502
  [0810.3588].

\bibitem{Lambda}
  S.~Aoki {\it et al.} [Flavour Lattice Averaging Group],
  {\it FLAG Review 2019,}
  1902.08191.

\bibitem{lat1a}
  C.~Allton {\it et al.},
  {\it 2+1 flavour thermal studies on an anisotropic lattice,}
  PoS LATTICE {2013} (2014) 151
  [1401.2116].

\bibitem{lat1b}
  G.~Aarts {\it et al}, 
  {\it The bottomonium spectrum at finite temperature 
  from N$_{f}$ = 2 + 1 lattice QCD,}
  JHEP {07} (2014) 097
  [1402.6210].

\bibitem{screening}
  B.B.~Brandt, A.~Francis, M.~Laine and H.B.~Meyer,
  {\it A relation between screening masses and real-time rates,}
  JHEP {05} (2014) 117
  [1404.2404].

\end{thebibliography}
\end{document}